\documentclass[aps,prd,twocolumn,nofootinbib,showpacs]{revtex4-1}

\usepackage{graphicx}

\usepackage[colorlinks=true, linkcolor=blue, citecolor=blue, urlcolor=blue]{hyperref}

\begin{document}

\title{Properties of the decay $H\to\gamma\gamma$ using the approximate $\alpha_s^4$-corrections and the principle of maximum
conformality}

\author{Qing Yu$^1$}
\email{yuq@cqu.edu.cn}
\author{Xing-Gang Wu$^1$}
\email{wuxg@cqu.edu.cn}
\author{Sheng-Quan Wang$^{2,3}$}
\email{sqwang@cqu.edu.cn}
\author{Xu-Dong Huang$^1$}
\email{hxud@cqu.eud.cn}
\author{Jian-Ming Shen$^1$}
\email{cqusjm@cqu.edu.cn}
\author{Jun Zeng$^1$}
\email{zengj@cqu.edu.cn}

\affiliation{$^1$ Department of Physics, Chongqing University, Chongqing 401331, People's Republic of China}
\affiliation{$^2$ SLAC National Accelerator Laboratory, Stanford University, Stanford, California 94039, USA}
\affiliation{$^3$ Department of Physics, Guizhou Minzu University, Guiyang 550025, People's Republic of China}

\date{\today}

\begin{abstract}

The Higgs boson decay channel, $H\to\gamma\gamma$, is one of the most important channels for probing the properties of the Higgs boson. In the paper, we reanalyze its decay width by using the QCD corrections up to $\alpha_s^4$-order level. The principle of maximum conformality has been adopted to achieve a precise pQCD prediction without conventional renormalization scheme-and-scale ambiguities. By taking the Higgs mass as the one given by the ATLAS and CMS collaborations, i.e. $M_{H}=125.09\pm0.21\pm0.11$ GeV, we obtain $\Gamma(H\to \gamma\gamma)|_{\rm LHC}=9.364^{+0.076}_{-0.075}$ KeV.


\end{abstract}

\maketitle

After the discovery of the Higgs boson at the Large Hadron Collider (LHC) in year 2012~\cite{Aad:2012tfa, Chatrchyan:2012xdj, ATLAS:2013mma, CMS:yva}, the remaining task is to confirm and learn more of its various properties either experimentally or theoretically. Theoretically, it is important to study its various decay modes within the framework of Standard Model (SM). As an important example, the Higgs decays into two photons, $H\to\gamma\gamma$, which could be observed at the LHC or a high luminosity $e^+ e^-$ linear collider, shall provide a clean platform for studying the Higgs properties.

The SM Higgs couples dominantly to the massive particles, the leading-order (LO) term of $H\to\gamma\gamma$ is already at the one-loop level, which inversely makes its higher-order QCD corrections very complex. At present, the LO, the next-to-leading order (NLO), the N$^2$LO, the approximate N$^3$LO, and the approximate N$^4$LO predictions on the decay width $\Gamma(H\to\gamma\gamma)$ have been done in Refs.\cite{Ellis:1975ap, Shifman:1979eb, Zheng:1990qa, Dawson:1992cy, Djouadi:1990aj, Djouadi:1993ji, Melnikov:1993tj, Inoue:1994jq, Spira:1995rr, Fleischer:2004vb, Harlander:2005rq, Aglietti:2006tp, Maierhofer:2012vv, Sturm:2014nva}. As shall be shown below, even though only the fermionic contributions which form a gauge-invariant subset have been considered in the N$^3$LO and N$^4$LO terms~\cite{Sturm:2014nva}, those state-of-art terms give us the opportunity to achieve a more precise prediction on $\Gamma(H\to\gamma\gamma)$.

According to the renormalization group invariance, the perturbatively calculable physical observable, corresponding to an infinite perturbative series, should be independent to any choices of the renormalization scheme and renormalization scale. Due to the mismatching of the running coupling ($\alpha_s$) and its coefficients at the same order, there is renormalization scheme-and-scale ambiguities for the fixed-order pQCD predictions. Those ambiguities always provide key errors for pQCD predictions, which are generally assumed to be decreased when more loop terms have been included. For example, Ref.\cite{Sturm:2014nva} shows that when going from the LO level to the approximate N$^4$LO level, the renormalization scale dependence decreases continuously. However, such decreasing of scale dependence is caused by compensation of scale dependence for all orders. Conventionally, the scale is chosen as the typical momentum flow the process, there are however many problems for such conventional treatment~\cite{Wu:2013ei, Wu:2014iba}. It is thus important to find a proper scale-setting way to set the renormalization scale so as to achieve an accurate fixed-order prediction.

In the literature, the principle of maximum conformality (PMC)~\cite{Brodsky:2011ta, Brodsky:2012rj, Brodsky:2011ig, Mojaza:2012mf, Brodsky:2013vpa} has been suggested to eliminate the renormalization scheme-and-scale ambiguities. Its key idea is to fix the renormalization scale based on the renormalization group equation (RGE); and when one applies the PMC, all non-conformal terms that govern the $\alpha_s$-running behavior of the pQCD approximant, should be systematically resummed. The PMC prediction satisfies the renormalization group invariance and all the self-consistency conditions of the renormalization group~\cite{Brodsky:2012ms}. Since the PMC resums all of the $\{\beta_i\}$-terms, the divergent renormalon terms like $n!\beta_0^n\alpha_s^n$ generally disappear, naturally leading to a convergent pQCD series. Furthermore, the definite PMC conformal series can also be adopted to reliably predict the contributions from uncalculated high-order terms~\cite{Du:2018dma}. In this paper we apply the PMC to set the renormalization scale for the decay width $\Gamma(H\to\gamma\gamma)$ up to N$^4$LO level and show that an accurate scale-independent prediction can indeed be achieved. For clarity, we shall adopt the PMC single-scale approach (PMC-s)~\cite{Shen:2017pdu} to do the scale-setting.

The decay width of the Higgs decays into two photons at the one-loop level takes the form
\begin{equation}
\Gamma(H \rightarrow \gamma\gamma) = \frac{M_{H}^{3}} {64\pi} \left|A_{W}+\sum\limits_{f} A_{f}\right|^{2}, \label{htolorr}
\end{equation}
where $M_H$ is the Higgs mass, $A_W$ denotes the contribution which arises from purely bosonic diagrams, and $A_f$ stands for the contribution from the amplitudes with $f=(t, b, c, \tau)$, which corresponds to top quark, bottom quark, charm quark and $\tau$ lepton, respectively.

The higher-order N$^2$LO, N$^3$LO and N$^4$LO expressions given in Refs.\cite{Maierhofer:2012vv, Sturm:2014nva} are for the top-quark running mass $(m_t)$. As has been argued in Ref.\cite{Wang:2013bla}, we have to transform those terms into the ones for the top-quark pole mass $(M_t)$ so as to avoid the entanglement of the $\{\beta_i\}$-terms from either the top-quark anomalous dimension or the RGE, thus avoiding the ambiguity in applying the PMC. Such transformation of mss can be done by using the relation between $m_t$ and $M_t$, and the relation up to ${\cal O}(\alpha_s^5)$ level can be read from Ref.\cite{Marquard:2016dcn}.

For convenience, we rewrite the decay width as
\begin{eqnarray}
\Gamma(H \rightarrow \gamma\gamma) &=& \frac{M_{H}^{3}} {64\pi} \bigg( A_{\rm LO}^{2} + A_{\rm EW} {\alpha\over \pi}\bigg) + R(\mu_r),
\label{hrrconv}
\end{eqnarray}
where $\alpha$ is the fine-structure constant. The LO contribution $A_{\rm LO}$ and the electroweak (EW) correction $A_{\rm EW}$ are~\cite{Maierhofer:2012vv}
\begin{eqnarray}
A_{\rm LO} &=& A_{W}^{(0)}+A_{f}^{(0)}+\hat A_t A_{t}^{(0)}, \\
A_{\rm EW} &=& 2 A_{\rm LO} A^{(1)}_{\rm EW} ,
\end{eqnarray}
where $A_W^{(0)}$ is the purely bosonic contribution to the amplitude, $A_f^{(0)}$ is the contribution from the amplitude with $f=(b,c,\tau)$, $\hat{A}_t = 2 Q_t^2 \alpha \sqrt{\sqrt{2} G_F}/\pi$, $G_F$ is the Fermi constant, and $Q_t$ is the top-quark electric charge. All of them have been calculated in Refs.\cite{Ellis:1975ap, Shifman:1979eb}, i.e.
\begin{eqnarray}
A_W^{(0)} &=& -\frac{\alpha\sqrt{\sqrt{2}G_F}}{2\pi} \left[ 2+\frac{3}{\tau_W}+\frac{3}{\tau_W}\left(2-{1\over\tau_W}\right) f(\tau_W) \right], \nonumber\\
A_f^{(0)} &=& \sum_{f=c,b,\tau} 3\frac{\alpha \sqrt{\sqrt{2} G_F}}{\pi \tau_{f}} Q_f^2 \left[1+\bigg(1-{1\over \tau_{f}}\bigg) f(\tau_f)\right], \nonumber\\
A_t^{(0)} &=& 1+{7\over30} \tau_t+{2\over21} \tau_t^2+{26\over525} \tau_t^3+{512\over17325} \tau_t^4 + {1216\over63063} \tau_t^5\nonumber\\
&& +{128\over9555} \tau_t^6, \nonumber
\end{eqnarray}
where
\begin{displaymath}
f(\tau)=\left\{
\begin{array}{ll}
{\rm Arcsin}^2(\sqrt{\tau}) &\quad \mbox{for}\quad \tau\leq 1\\
-\frac{1}{4}\left[\ln\frac{1+\sqrt{1-\tau^{-1}}}{1-\sqrt{1-\tau^{-1}}}-i\pi\right]^2&
\quad \mbox{for}\quad \tau>1
\end{array} \right. ,
\end{displaymath}
$Q_f$ denotes the electric charge for $f=(c,b,\tau)$, $\tau_W=M^2_H/(4M_W^2)$, $\tau_t = M_H^2/(4 M_t^2)$ and $\tau_{f}=M^2_H/(4 M_f^2)$, and the expression for the NLO electroweak term $A^{(1)}_{\rm EW}$ can be read from Refs.\cite{Marquard:2016dcn, Actis:2008ts}.

The QCD corrections to the decay width $\Gamma(H \to \gamma\gamma) $ are separately represented by $R(\mu_r)$, whose perturbative series up to $(n+1)$-loop level can be written as
\begin{equation}
R_n(\mu_r) = \sum^{n}_{i=1} r_{i}(\mu) a_s^{i}(\mu_r),
\end{equation}
where $a_s={\alpha_s}/\pi$, $\mu_r$ is the renormalization scale. The perturbative coefficients $r_i$ in the $\overline{\rm MS}$-scheme up to $\alpha_s^4$-order level can be derived from Refs.~\cite{Maierhofer:2012vv, Sturm:2014nva}. To apply the PMC, the $n_f$-power series ($n_f$ being the active flavor number) in the coefficients $r_i$ should be rewritten into conformal terms and non-conformal $\beta_i$-terms~\cite{Mojaza:2012mf, Brodsky:2013vpa}, i.e.
\begin{eqnarray}
r_1 &=& r_{1,0},  \\
r_2 &=& r_{2,0} + r_{2,1} \beta_0,  \\
r_3 &=& r_{3,0} + r_{2,1} \beta_1 + 2r_{3,1} \beta_0 + r_{3,2} \beta_0^2,  \\
r_4 &=& r_{4,0} + r_{2,1} \beta_2 + 2r_{3,1} \beta_1 + \frac{5}{2} r_{3,2} \beta_0 \beta_1  \\
&& + 3r_{4,1} \beta_0 + 3r_{4,2} \beta_0^2 + r_{4,3} \beta_0^3, \\
&& \!\!\!\!\!\!\!\!\!\!\!\! \cdots,
\end{eqnarray}
where the $\beta$-pattern at each perturbative order is a superposition of RGE, and all the coefficients $r_{i,j}$ can be fixed from the $n_f$-power series at the same order by using the degeneracy relations among different orders. $r_{i,0}$ are conformal coefficients which are exactly free of $\mu_r$ for the present channel, and $r_{i,j(j\neq 0)}$ are non-conformal coefficients which are functions of $\mu_r$, i.e.,
\begin{equation}
r_{i,j} = \sum_{k=0}^{j} C_j^k \hat{r}_{i-k,j-k} \ln^k(\mu_r^2/M_H^2),
\end{equation}
where $\hat{r}_{i,j}=r_{i,j}|_{\mu_r=M_H}$. The needed $\{\beta_i\}$-functions also under the $\overline{\rm MS}$-scheme are available in Refs.\cite{Gross:1973id, Politzer:1973fx, Caswell:1974gg, Tarasov:1980au, Larin:1993tp, vanRitbergen:1997va, Chetyrkin:2004mf, Czakon:2004bu, Baikov:2016tgj}.

Following the standard procedures of the PMC single-scale approach~\cite{Shen:2017pdu}, the pQCD corrections to the decay width $\Gamma(H \to \gamma\gamma)$ can be simplified as
\begin{equation}
R_n(\mu_r)|_{\rm PMC} = \sum^{n}_{i=1} \hat{r}_{i,0} a_s^i(Q_\star),
\label{htorrpmc}
\end{equation}
where $Q_{\star}$ is the PMC scale. Using the known pQCD corrections up to N$^4$LO level, $Q_{\star}$ can be fixed up to next-to-next-to-leading-log ($\rm N^2LL$) accuracy, i.e.,
\begin{equation}
\ln\frac{Q_\star^2}{M_H^2} = \sum_{i} T_{i} a^{i}_s(M_H),
\label{htorrpmcscale}
\end{equation}
whose first three coefficients with $i=(0,1,2)$ can be determined by the known five-loop QCD corrections to the decay width $\Gamma(H \to \gamma\gamma)$, which are
\begin{eqnarray}
T_0 &=& -{\hat{r}_{2,1}\over \hat{r}_{1,0}},  \\
T_1 &=& {2(\hat{r}_{2,0}\hat{r}_{2,1}-\hat{r}_{1,0}\hat{r}_{3,1})\over \hat{r}_{1,0}^2} +{(\hat{r}_{2,1}^2-\hat{r}_{1,0}\hat{r}_{3,2})\over \hat{r}_{1,0}^2}\beta_0,  \\
T_2 &=&{4(\hat{r}_{1,0}\hat{r}_{2,0}\hat{r}_{3,1}-\hat{r}_{2,0}^2\hat{r}_{2,1}) +3(\hat{r}_{1,0}\hat{r}_{2,1}\hat{r}_{3,0}-\hat{r}_{1,0}^2\hat{r}_{4,1})\over \hat{r}_{1,0}^3 }\nonumber\\
&-&{\hat{r}_{2,0}\hat{r}_{2,1}^2 +2(\hat{r}_{2,0}\hat{r}_{2,1}^2-2\hat{r}_{1,0}\hat{r}_{2,1}\hat{r}_{3,1} -\hat{r}_{1,0}\hat{r}_{2,0}\hat{r}_{3,2})\over \hat{r}_{1,0}^3}\beta_0\nonumber\\
&-&{3\hat{r}_{1,0}^2\hat{r}_{4,2}\over \hat{r}_{1,0}^3}\beta_0 +{3(\hat{r}_{2,1}^2-\hat{r}_{1,0}\hat{r}_{3,2})\over 2\hat{r}_{1,0}^2}\beta_1 \nonumber\\
&+&{(\hat{r}_{1,0}\hat{r}_{2,1}\hat{r}_{3,2}-\hat{r}_{1,0}^2\hat{r}_{4,3})+(\hat{r}_{1,0}\hat{r}_{2,1}\hat{r}_{3,2}-\hat{r}_{2,1}^3)\over \hat{r}_{1,0}^3}\beta_0^2.
\end{eqnarray}
Eq.(\ref{htorrpmcscale}) indicates that the scale $Q_*$ is free of $\mu_r$, together with the fact that the conformal coefficients $\hat{r}_{i,0}$ are also free of $\mu_r$, the net PMC prediction $R_n(\mu_r)|_{\rm PMC}$ is scale-independent; Thus the conventional scale ambiguity is eliminated. As a subtle point, due to unknown even higher-order terms in $Q_*$ perturbative series, there is residual scale dependence for $Q_*$. However such kind of residual scale dependence is different from conventional renormalization scale ambiguity, which is usually negligible due to both the $\alpha_s$-suppression and the exponential suppression. This property has been confirmed in many PMC applications done in the literature.

To do the calculation, we take the following ones as their central values~\cite{Tanabashi:2018oca}: the $W$-boson mass $M_{W}=80.379$ GeV, the $\tau$-lepton mass $M_{\tau}=1.77686$ GeV, the $b$-quark pole mass $M_{b}=4.78$ GeV, the $c$-quark pole mass $M_{c}=1.67$ GeV, the $t$-quark pole mass $M_{t}=173.07$ GeV, and the Higgs mass $M_{H}=125.9$ GeV. The Fermi constant $G_{F}=1.1663787\times10^{-5}$ $\rm GeV^{-2}$ and the fine structure constant $\alpha=1/137.035999139$. We adopt the four-loop $\alpha_s$-running and $\alpha_s(M_{Z}=91.1876{\rm GeV})=0.1181$ to fix the $\alpha_{s}$-running behavior.

\begin{figure}[htb]
\centering
\includegraphics[width=0.48\textwidth]{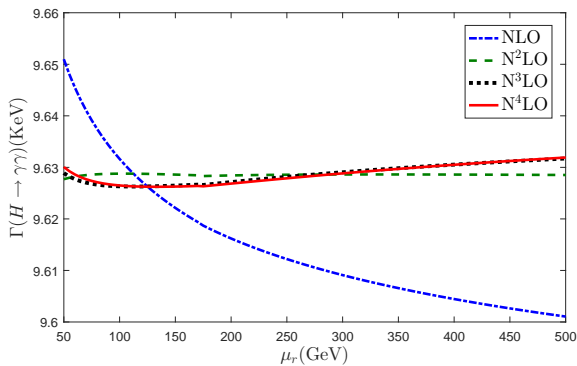}
\caption{Total decay width $H\to\gamma\gamma$ versus the initial scale $\mu_r$ up to N$^{4}$LO level under conventional scale-setting.}
\label{fig:convDecayWidth}
\end{figure}

\begin{figure}[htb]
\centering
\includegraphics[width=0.48\textwidth]{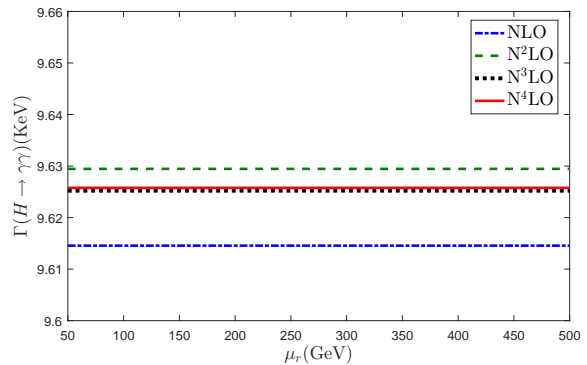}
\caption{Total decay width $H\to\gamma\gamma$ versus the initial scale $\mu_r$ up to N$^{4}$LO level under PMC scale-setting.}
\label{fig:pmcDecayWidth}
\end{figure}

Firstly, we present the total decay width $\Gamma(H\to\gamma\gamma)$ up to N$^4$LO level under conventional and PMC scale-settings in Figs.(\ref{fig:convDecayWidth}, \ref{fig:pmcDecayWidth}). Fig.(\ref{fig:convDecayWidth}) shows that under conventional scale-setting, the scale dependence becomes smaller when more loop terms are included. This agrees with the conventional wisdom that by finishing enough higher-order calculation, one may achieve a desirable scale-independent prediction. The N$^4$LO total decay width under conventional scale-setting gives
\begin{eqnarray}
\Gamma(H\to\gamma\gamma)|_{\rm Conv.} &=& 9.626^{+0.002}_{+0.002}\; {\rm KeV},
\end{eqnarray}
where central value is for $\mu_r=M_H$, and the renormalization scale error is for $\mu_r\in[M_{H}/2, 2M_{H}]$.

\begin{table*}[tb]
\centering
\caption{The decay width $\Gamma(H\to\gamma\gamma)$ under conventional (Conv.) and PMC scale-settings. $\Gamma_{\rm LO+EW}$, $\Gamma_{\rm NLO}$, $\Gamma_{\rm N^{2}LO}$, $\Gamma_{\rm N^{3}LO}$ and $\Gamma_{\rm N^{4}LO}$ are individual decay widths at LO+EW, NLO, N$^{2}$LO, N$^{3}$LO and N$^{4}$LO levels, respectively. $\Gamma_{\rm Total}$ is the total decay width up to N$^{4}$LO level, including the electroweak correction $\Gamma_{\rm EW}$. Three typical scales $\mu_{r}$ = $M_{H}/2$, $M_{H}$, $2 M_{H}$ are adopted.} \label{Total-pmcs}
\begin{tabular}{cccccccc}
\hline
~~~~&~~ ~~&~~$i={\rm LO+EW}$~~&~~$i={\rm NLO}$~~&~~$i={\rm N^2LO}$~~&~~${\rm N^3LO}$~~&~~$i={\rm N^4LO}$~~&~~$i={\rm Total}$~~ \\
\hline
               & $\mu_r=M_H/2$            & $9.46477$  & $0.17927$ & $-0.01573$ & $-0.00085$ & $ 0.00083$ & $9.62830$   \\
${\Gamma_i \rm{(KeV)}}|_{\rm Conv.}$      & $\mu_r=M_H$    & $9.46477$ & $0.16133$ & $ 0.00263$ & $-0.00242$ & $-0.00007$ & $9.62624$   \\
              & $\mu_r=2M_H$              & $9.46477$ & $0.14731$ & $ 0.01649$ &  $-0.00038$ & $-0.00028$ & $9.62791$   \\
\hline
${\Gamma_i\rm{(KeV)}}|_{\rm PMC}$  & $\mu_r\in[M_H/2,2M_H]$ & $9.46477$ & $0.14979$ & $ 0.01489$ &  $-0.00423$ & $ 0.00056$ & $9.62578$  \\
\hline
\end{tabular}
\end{table*}

It is noted that such nearly scale-independence for the N$^4$LO total decay width $\Gamma(H\to\gamma\gamma)$ under conventional scale-setting is caused by large cancellations of the scale dependence among different orders. This can be explicitly seen from Table~\ref{Total-pmcs}, in which the individual decay widths at LO+EW, NLO, N$^{2}$LO, N$^{3}$LO and N$^{4}$LO levels are presented. We define a parameter $\kappa_i$ to measure the scale dependence of the separate decay widths at different orders, i.e.
\begin{equation}
\kappa_i=\left|\frac{\left. \Gamma_{i}\right|_{\mu_{r}=M_H/2} -\left. \Gamma_{i}\right|_{\mu_{r}=2M_H}}{\left. \Gamma_{i} \right|_{\mu_{r}=M_H}}\right|,
\end{equation}
where the subscript $i$ stands for NLO, N$^2$LO, N$^3$LO, N$^4$LO and Total decay widths, respectively. Table~\ref{Total-pmcs} shows that under conventional scale-setting,
\begin{eqnarray}
\kappa_{\rm NLO}        &=& 20\%, \;\;\; \kappa_{\rm {N^2LO}}  = 1.2\times10^3\%, \\
\kappa_{\rm {N^3LO}} &=& 19\%,  \;\;\; \kappa_{\rm {N^4LO}} = 1.6\times10^3\%.
\end{eqnarray}
Large magnitude of $\kappa$ indicates that under conventional scale-setting, there are large scale errors for each orders, and such kind of scale errors cannot be affected by the high-order terms.

On the other hand, as shown by Fig.(\ref{fig:pmcDecayWidth}), the PMC prediction is almost scale-independent for each order, and the PMC prediction on $\Gamma(H\to\gamma\gamma)$ quickly approaches its ``physical'' value due to a faster convergence than conventional pQCD series. Because the magnitude of the newly added N$^3$LO and N$^4$LO terms are only about $28\%$ and $4\%$ of that of the N$^2$LO terms whose magnitude is small, our previous N$^2$LO PMC prediction agrees with the present prediction~\cite{Wang:2013akk} with high precision. Table~\ref{Total-pmcs} shows that after applying PMC scale-setting, both the separate decay widths and the total decay width are nearly unchanged for $\mu_r\in[M_{H}/2, 2M_{H}]$. The N$^4$LO total decay width under PMC scale-setting is
\begin{equation}
\Gamma(H\to\gamma\gamma)|_{\rm PMC} \equiv 9.626\; {\rm KeV}.
\end{equation}

\begin{figure}[htb]
\includegraphics[width=0.48\textwidth]{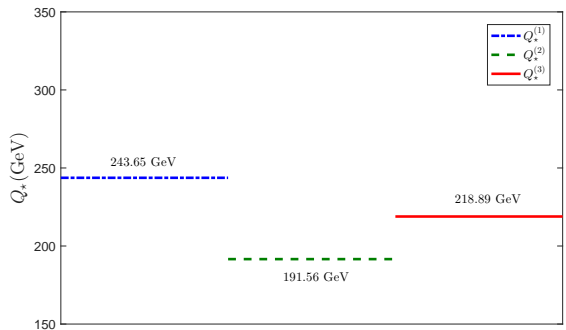}
\caption{The determined effective scale $Q_\star$. $Q^{(1)}_\star$ is at the LL accuracy, $Q^{(2)}_\star$ is at the NLL accuracy and $Q^{(3)}_\star$ is at the ${\rm N^{2}LL}$ accuracy.}
\label{fig:pmcscale}
\end{figure}

The four-loop and five-loop fermionic contributions are helpful to set an accurate PMC scale. The nearly scale-independence for each order under PMC scale-setting is caused by the fact that the effective PMC scale $Q_\star$ can be fixed up to ${\rm N^{2}LL}$-accuracy by using the known five-loop pQCD corrections, i.e.
\begin{equation}
\ln\frac{Q_\star^2}{M_H^2}=1.321-4.271\alpha_s(M_H)+21.029\alpha^2_s(M_H) .
\end{equation}
We present $Q_\star$ in Fig.(\ref{fig:pmcscale}), in which $Q^{(1)}_\star$ is computed at the LL accuracy, $Q^{(2)}_\star$ is at the NLL accuracy and $Q^{(3)}_\star$ is at the ${\rm N^{2}LL}$ accuracy, respectively. Fig.(\ref{fig:pmcscale}) shows that similar to the case of $H\to b\bar{b}$~\cite{Shen:2017pdu}, $Q_\star$ oscillates as more loop corrections are included, which shall approach its accurate value as more loop terms are included.

\begin{figure}[htb]
\centering
\includegraphics[width=0.45\textwidth]{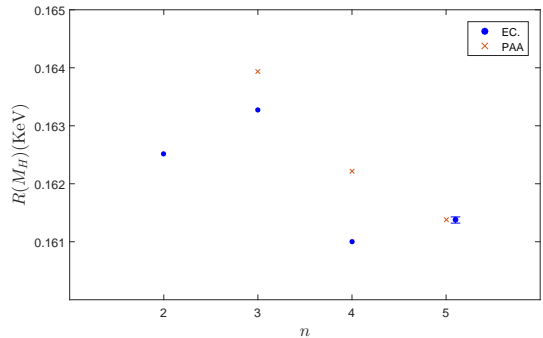}
\caption{Comparison of the exact (``EC'') and the predicted ([$0/(n-1)$]-type ``PAA'') pQCD approximant $R_{n}(M_H)$ under PMC scale-setting. It shows how the PAA predictions change when more loop-terms are included. }
\label{Fig:Hgammagamma}
\end{figure}

Secondly, the Pad$\acute{\rm{e}}$ approximation approach (PAA) provides a practical way for promoting a finite series to an analytic function~\cite{Basdevant:1972fe, Samuel:1992qg, Samuel:1995jc}, which has recently been suggested to give a reliable prediction of uncalculated high-order terms by using the PMC conformal series~\cite{Du:2018dma}. As an attempt, following the same method described in detail in Ref.\cite{Du:2018dma}, we give a PAA+PMC prediction for $R_{n}(M_H)$ by using the preferable $[0/(n-1)]$-type Pad\'e series, which is in Fig.(\ref{Fig:Hgammagamma}). The predicted results decrease rapidly as more high-order loop terms are presented, e.g., the exact (``EC'') and the predicted $R_{n}(M_H)$ tend to be stable as more higher loops are absorbed. The difference for $R_{3,4}(M_H)|_{\rm PAA}$ and $R_{3,4}(M_H)|_{\rm EC}$ is already less than $1\%$, thus the exact value of $R(M_H)|_{\rm EC}$ could be directly teated as $R_{5}(M_H)|_{\rm PAA}$, which gives
\begin{equation}
R_5(M_H) \cong 1.614\times10^{-1}\; {\rm KeV}.
\label{Hyy4loop}
\end{equation}
Then the total decay width
\begin{equation}
\Gamma_5(H\to\gamma\gamma)|_{\rm PMC} = \left[9.626\pm5.354\times10^{-5}\right]\; {\rm KeV},
\end{equation}
where the error is the PAA+PMC prediction of uncalculated high-order pQCD contributions, which is negligible.

\begin{figure}[htb]
\centering
\includegraphics[width=0.48\textwidth]{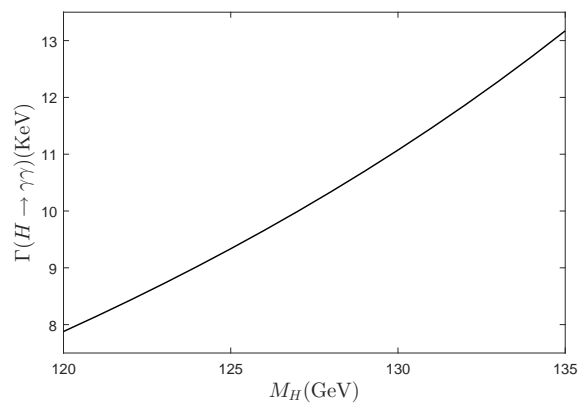}
\caption{The PMC prediction of the decay width $\Gamma(H\to\gamma\gamma)$ versus the Higgs mass $M_{H}$. } \label{Plot:sqpms:c}
\end{figure}

The total decay width $\Gamma(H\to\gamma\gamma)$ versus the Higgs mass $M_{H}$ is presented in Fig.(\ref{Plot:sqpms:c}). If $M_H=125.9$ GeV, our calculation shows that $9.626$ KeV could be the ``exact" value for the total decay width of $H\to\gamma\gamma$. If taking the Higgs mass as the one given by the ATLAS and CMS collaborations~\cite{Aad:2015zhl, Mei:2017eev}, i.e. $M_{H}=125.09\pm0.21\pm0.11$ GeV, we obtain
\begin{equation}
\Gamma(H\to \gamma\gamma)|_{\rm LHC} = 9.364^{+0.076}_{-0.075} \; {\rm KeV},
\end{equation}

\begin{figure}[htb]
\centering
\includegraphics[width=0.5\textwidth]{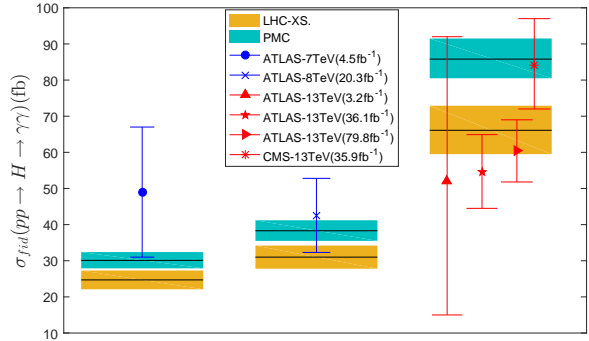}
\caption{The fiducial cross section $\sigma_{\rm fid}(pp\to H\to \gamma\gamma)$ using the $\Gamma(H\to\gamma\gamma)$ up to N$^4$LO level. The LHC-XS prediction~\cite{Heinemeyer:2013tqa}, the ATLAS measurements~\cite{TheATLAScollaboration:2015poe, ATLAS:2017myr, ATLAS:2018uso} and the CMS measurement~\cite{CMS:2017nyv} are presented as a comparison. }
\label{sigmafidATLAS7813}
\end{figure}

Thirdly, as an application of the $H\to\gamma\gamma$ decay width, we predict the ``fiducial cross section" of the process $pp\to H\to\gamma\gamma$, which has been predicted by the LHC-XS group under the conventional scale-setting~\cite{Heinemeyer:2013tqa} and has been measured by ATLAS and CMS collaborations with increasing integrated luminocities~\cite{TheATLAScollaboration:2015poe, ATLAS:2017myr, ATLAS:2018uso, CMS:2017nyv}. A PMC prediction has previously been given in Ref.\cite{Wang:2016wgw} by using $\Gamma(H\to\gamma\gamma)$ up to N$^2$LO level. Taking the same parameters as those of Refs.\cite{Heinemeyer:2013tqa, Wang:2016wgw, deFlorian:2016spz}, e.g. $M_H=125$ GeV and $M_t$=173.3 GeV, and by using the present $\Gamma(H\to\gamma\gamma)$ up to N$^4$LO level, we obtain $\sigma_{\rm fid}(pp\to H\to \gamma\gamma)=30.1^{+2.3}_{-2.2}$ fb, $38.3^{+2.9}_{-2.8}$ fb, and $85.8^{+5.7}_{-5.3}$ fb for the proton-proton center-of-mass collision energy $\sqrt{S}=7$, $8$ and $13$ TeV, respectively. Here the
errors are dominated by the error of the Higgs inclusive cross section. A comparison of the recent experimental data is put in Fig.(\ref{sigmafidATLAS7813}). A better agreement with the data at $\sqrt{S}=7$, $8$ TeV can be achieved by applying the PMC. The ATLAS and CMS measurements at $\sqrt{S}=13$ TeV are still of large errors and are in disagreement, and the PMC prediction prefers the CMS data~\cite{CMS:2017nyv}.

As a summary, the PMC uses the basic RGE to set the $\alpha_s$-running behavior; the resultant conformal series is independent to the initial choice of renormalization scale and renormalization scheme, and thus eliminates the conventional scheme-and-scale ambiguities. By using the known QCD corrections up to the known approximate five-loop level, we can fix the effective PMC scale up to N$^2$LL level and an accurate scheme-and-scale independent prediction for the decay width $\Gamma(H\to\gamma\gamma)$ can be achieved. The residual scale dependence due to unknown high-order terms are negligible, and the PMC prediction is almost scale-independent for each order. The PAA+PMC treatment indicates that our present N$^4$LO decay width $\Gamma(H\to\gamma\gamma)|_{\rm PMC}$ is already approaches its ``physical'' value, since the contribution of the uncalculated even higher-order is negligible due to a more convergent renormalon-free conformal series.

\hspace{1cm}

{\bf Acknowledgments}: This work was supported in part by Natural Science Foundation of China under Grant No.11625520, No.11547010 and No.11705033; by the Project of Guizhou Provincial Department of Science and Technology under Grant No.2016GZ42963 and the Key Project for Innovation Research Groups of Guizhou Provincial Department of Education under Grant No.KY[2016]028 and No.KY[2017]067.


\begin{thebibliography}{99}

\bibitem{Aad:2012tfa}
  G.~Aad {\it et al.} [ATLAS Collaboration],
  Phys.\ Lett.\ B {\bf 716}, 1 (2012).

\bibitem{Chatrchyan:2012xdj}
  S.~Chatrchyan {\it et al.} [CMS Collaboration],
  Phys.\ Lett.\ B {\bf 716}, 30 (2012).

\bibitem{ATLAS:2013mma}
  The ATLAS Collaboration [ATLAS Collaboration],
  ATLAS-CONF-2013-014.

\bibitem{CMS:yva}
  The CMS Collaboration [CMS Collaboration],
  CMS-PAS-HIG-13-005.

\bibitem{Ellis:1975ap}
  J.~R.~Ellis, M.~K.~Gaillard and D.~V.~Nanopoulos,
  Nucl.\ Phys.\ B {\bf 106}, 292 (1976).

\bibitem{Shifman:1979eb}
  M.~A.~Shifman, A.~I.~Vainshtein, M.~B.~Voloshin and V.~I.~Zakharov,
  Sov.\ J.\ Nucl.\ Phys.\  {\bf 30}, 711 (1979).

\bibitem{Zheng:1990qa}
  H.~Q.~Zheng and D.~D.~Wu,
  Phys.\ Rev.\ D {\bf 42}, 3760 (1990).

\bibitem{Dawson:1992cy}
  S.~Dawson and R.~P.~Kauffman,
  Phys.\ Rev.\ D {\bf 47}, 1264 (1993).

\bibitem{Djouadi:1990aj}
  A.~Djouadi, M.~Spira, J.~J.~van der Bij and P.~M.~Zerwas,
  Phys.\ Lett.\ B {\bf 257}, 187 (1991).

\bibitem{Djouadi:1993ji}
  A.~Djouadi, M.~Spira and P.~M.~Zerwas,
  Phys.\ Lett.\ B {\bf 311}, 255 (1993).

\bibitem{Melnikov:1993tj}
  K.~Melnikov and O.~I.~Yakovlev,
  Phys.\ Lett.\ B {\bf 312}, 179 (1993).

\bibitem{Inoue:1994jq}
  M.~Inoue, R.~Najima, T.~Oka and J.~Saito,
  Mod.\ Phys.\ Lett.\ A {\bf 9}, 1189 (1994).

\bibitem{Spira:1995rr}
  M.~Spira, A.~Djouadi, D.~Graudenz and P.~M.~Zerwas,
  Nucl.\ Phys.\ B {\bf 453}, 17 (1995).

\bibitem{Fleischer:2004vb}
  J.~Fleischer, O.~V.~Tarasov and V.~O.~Tarasov,
  Phys.\ Lett.\ B {\bf 584}, 294 (2004).

\bibitem{Harlander:2005rq}
  R.~Harlander and P.~Kant,
  JHEP {\bf 0512}, 015 (2005).

\bibitem{Aglietti:2006tp}
  U.~Aglietti, R.~Bonciani, G.~Degrassi and A.~Vicini,
  JHEP {\bf 0701}, 021 (2007).

\bibitem{Maierhofer:2012vv}
  P.~Maierhofer and P.~Marquard,
  Phys.\ Lett.\ B {\bf 721}, 131 (2013).

\bibitem{Sturm:2014nva}
  C.~Sturm,
  Eur.\ Phys.\ J.\ C {\bf 74}, 2978 (2014).

\bibitem{Wu:2013ei}
  X.~G.~Wu, S.~J.~Brodsky and M.~Mojaza,
  Prog.\ Part.\ Nucl.\ Phys.\  {\bf 72}, 44 (2013).

\bibitem{Wu:2014iba}
  X.~G.~Wu, Y.~Ma, S.~Q.~Wang, H.~B.~Fu, H.~H.~Ma, S.~J.~Brodsky and M.~Mojaza,
  Rep.\ Prog.\ Phys.\  {\bf 78}, 126201 (2015).

\bibitem{Brodsky:2011ta}
  S.~J.~Brodsky and X.~G.~Wu,
  Phys.\ Rev.\ D {\bf 85}, 034038 (2012).

\bibitem{Brodsky:2012rj}
  S.~J.~Brodsky and X.~G.~Wu,
  Phys.\ Rev.\ Lett.\ {\bf 109}, 042002 (2012).

\bibitem{Brodsky:2011ig}
  S.~J.~Brodsky and L.~Di Giustino,
  Phys.\ Rev.\ D {\bf 86}, 085026 (2012).

\bibitem{Mojaza:2012mf}
  M.~Mojaza, S.~J.~Brodsky and X.~G.~Wu,
  Phys.\ Rev.\ Lett.\ {\bf 110}, 192001 (2013).

\bibitem{Brodsky:2013vpa}
  S.~J.~Brodsky, M.~Mojaza and X.~G.~Wu,
  Phys.\ Rev.\ D {\bf 89}, 014027 (2014).

\bibitem{Brodsky:2012ms}
  S.~J.~Brodsky and X.~G.~Wu,
  Phys.\ Rev.\ D {\bf 86}, 054018 (2012).

\bibitem{Du:2018dma}
  B.~L.~Du, X.~G.~Wu, J.~M.~Shen and S.~J.~Brodsky,
  arXiv:1807.11144 [hep-ph].

\bibitem{Shen:2017pdu}
  J.~M.~Shen, X.~G.~Wu, B.~L.~Du and S.~J.~Brodsky,
  Phys.\ Rev.\ D {\bf 95}, 094006 (2017).

\bibitem{Wang:2013bla}
  S.~Q.~Wang, X.~G.~Wu, X.~C.~Zheng, J.~M.~Shen and Q.~L.~Zhang,
  Eur.\ Phys.\ J.\ C {\bf 74}, 2825 (2014).

\bibitem{Marquard:2016dcn}
  P.~Marquard, A.~V.~Smirnov, V.~A.~Smirnov, M.~Steinhauser and D.~Wellmann,
  Phys.\ Rev.\ D {\bf 94}, 074025 (2016).

\bibitem{Actis:2008ts}
  S.~Actis, G.~Passarino, C.~Sturm and S.~Uccirati,
  Nucl.\ Phys.\ B {\bf 811}, 182 (2009).

\bibitem{Gross:1973id}
  D.~J.~Gross and F.~Wilczek,
  Phys.\ Rev.\ Lett.\  {\bf 30}, 1343 (1973).

\bibitem{Politzer:1973fx}
  H.~D.~Politzer,
  Phys.\ Rev.\ Lett.\  {\bf 30}, 1346 (1973).

\bibitem{Caswell:1974gg}
  W.~E.~Caswell,
  Phys.\ Rev.\ Lett.\  {\bf 33}, 244 (1974).

\bibitem{Tarasov:1980au}
  O.~V.~Tarasov, A.~A.~Vladimirov and A.~Y.~Zharkov,
  Phys.\ Lett.\  B {\bf 93}, 429 (1980).

\bibitem{Larin:1993tp}
  S.~A.~Larin and J.~A.~M.~Vermaseren,
  Phys.\ Lett.\ B {\bf 303}, 334 (1993).

\bibitem{vanRitbergen:1997va}
  T.~van Ritbergen, J.~A.~M.~Vermaseren and S.~A.~Larin,
  Phys.\ Lett.\ B {\bf 400}, 379 (1997).

\bibitem{Chetyrkin:2004mf}
  K.~G.~Chetyrkin,
  Nucl.\ Phys.\ B {\bf 710}, 499 (2005).

\bibitem{Czakon:2004bu}
  M.~Czakon,
  Nucl.\ Phys.\ B {\bf 710}, 485 (2005).

\bibitem{Baikov:2016tgj}
  P.~A.~Baikov, K.~G.~Chetyrkin and J.~H.~K¨¹hn,
  Phys.\ Rev.\ Lett.\  {\bf 118}, 082002 (2017).

\bibitem{Tanabashi:2018oca}
  M.~Tanabashi {\it et al.} [Particle Data Group],
  Phys.\ Rev.\ D {\bf 98}, 030001 (2018).

\bibitem{Wang:2013akk}
  S.~Q.~Wang, X.~G.~Wu, X.~C.~Zheng, G.~Chen and J.~M.~Shen,
  J.\ Phys.\ G {\bf 41}, 075010 (2014)

\bibitem{Basdevant:1972fe}
  J.~L.~Basdevant,
  Fortsch.\ Phys.\ {\bf 20}, 283 (1972).

\bibitem{Samuel:1992qg}
  M.~A.~Samuel, G.~Li and E.~Steinfelds,
  Phys.\ Lett.\ B {\bf 323}, 188 (1994).

\bibitem{Samuel:1995jc}
  M.~A.~Samuel, J.~R.~Ellis and M.~Karliner,
  Phys.\ Rev.\ Lett.\ {\bf 74}, 4380 (1995).

\bibitem{Aad:2015zhl}
  G.~Aad {\it et al.} [ATLAS and CMS Collaborations],
  Phys.\ Rev.\ Lett.\ {\bf 114}, 191803 (2015).

\bibitem{Mei:2017eev}
  H.~Mei [CMS Collaboration],
  CMS-CR-2017-123.

\bibitem{Heinemeyer:2013tqa}
  S.~Heinemeyer {\it et al.} [LHC Higgs Cross Section Working Group],
  CERN-2013-004.

\bibitem{TheATLAScollaboration:2015poe}
  G. Aad {\it et al.}, [ATLAS Collaboration],
  ATLAS-CONF-2015-060.

\bibitem{ATLAS:2017myr}
  The ATLAS collaboration [ATLAS Collaboration],
 ATLAS-CONF-2017-045.

\bibitem{ATLAS:2018uso}
  The ATLAS collaboration [ATLAS Collaboration],
  ATLAS-CONF-2018-028.

\bibitem{CMS:2017nyv}
  The CMS Collaboration [CMS Collaboration],
  CMS-PAS-HIG-17-015.

\bibitem{Wang:2016wgw}
  S.~Q.~Wang, X.~G.~Wu, S.~J.~Brodsky and M.~Mojaza,
  Phys.\ Rev.\ D {\bf 94}, 053003 (2016).

\bibitem{deFlorian:2016spz}
  D.~de Florian {\it et al.} [LHC Higgs Cross Section Working Group],
  CERN-2017-002-M.

\end{thebibliography}
\end{document}